\documentclass[11pt]{JHEP3}
\usepackage{cite,amsmath,amssymb,bbm}

\usepackage{booktabs}
\usepackage{colonequals}
\usepackage{accents}

\def\openone{\mathbbm{1}}

\def\dirac{\gamma_\mu p^\mu}

\def\p{\mbox{\boldmath$\displaystyle\mathbf{p}$}}
\def\k{\mbox{\boldmath$\displaystyle\mathbf{\kappa}$}}

\def\0{\mbox{\boldmath$\displaystyle\mathbf{0}$}}
\def\s{\mbox{\boldmath$\displaystyle\mathbf{\sigma}$}}
\def\z{\mbox{\boldmath$\displaystyle\mathbf{\zeta}$}}
\def\J{\mbox{\boldmath$\displaystyle\mathbf{J}$}}
\def\K{\mbox{\boldmath$\displaystyle\mathbf{K}$}}
\def\x{\mbox{\boldmath$\displaystyle\mathbf{x}$}}

\def\beq{\begin{equation}}
\def\eeq{\end{equation}}
\def\beqar{\begin{eqnarray}}
\def\eeqar{\end{eqnarray}}

\def\ce{\colonequals}

\def\dual#1{\accentset{\boldsymbol{\neg}\vspace{-.13ex}}{#1}}
\newcommand{\ud}{\,\mathrm{d}}

\title{Very special relativity as relativity of dark matter: the Elko connection}

\author{D. V. Ahluwalia and S. P. Horvath \\
Department of Physics and Astronomy, Rutherford Building \\
University of Canterbury, Private Bag 4800, Christchurch 8020, New Zealand \\
E-mail: \email{dharamvir.ahluwalia@canterbury.ac.nz, \\ sebastian.horvath@pg.canterbury.ac.nz}}

\abstract{In the \emph{very special relativity} (VSR) proposal by Cohen and Glashow, it was pointed out that invariance under \emph{HOM}(2) is both necessary and sufficient to explain the null result of the Michelson-Morely experiment. It is the quantum field theoretic demand of locality, or the requirement of \emph{P, T, CP}, or \emph{CT} invariance, that makes invariance under the Lorentz group a necessity. Originally it was conjectured that VSR operates at the Planck scale; we propose that the natural arena for VSR is at energies similar to the standard model, but in the dark sector. To this end we provide an \emph{ab initio} spinor representation invariant under the \emph{SIM}(2) avatar of VSR and construct a mass dimension one fermionic quantum field of spin one half. This field turns out to be a very close sibling of Elko and it exhibits the same striking property of intrinsic darkness with respect to the standard model fields. In the new construct, the tension between Elko and Lorentz symmetries is fully resolved. We thus entertain the possibility that the symmetries underlying the standard model matter and gauge fields are those of Lorentz, while the event space underlying the dark matter and the dark gauge fields supports the algebraic structure underlying VSR.
\vskip 21pt
\noindent{J\sc ournal Reference: JHEP11(2010)078}
}


\preprint{WestportAutumn2010 }
  
\begin{document}

\maketitle

\section{Introduction}  \label{Sec:intro}

Contrary to what has been historically assumed, Cohen and Glashow have shown that time dilation, the law of velocity addition, and the universal and isotropic velocity, do not require the full Poincar\'e group but can all be accounted for by two of the very special relativity subgroups \cite{Cohen:2006ky}. For this reason they ask us to entertain the ``possibility that the many empirical successes of special relativity need not demand Lorentz invariance of the underlying framework.'' If none of the discrete symmetries of $P$, $T$, $CP$ or $CT$ is violated, and a quantum field theoretic theory contains massive particles, then the choice is unique; the symmetries of the theory must be those of the Poincar\'e group. However, if any one of these discrete symmetries is violated, then it suffices that the symmetry group underlying the theory is isomorphic to one of two specific VSR subgroups. The largest of these subgroups is obtained by adjoining the four spacetime translation generators to a 4-parameter subgroup that is, up to isomorphism,~\emph{SIM}(2). Since $CP$ violating effects for the standard model (SM) sector are small one may then be tempted to consider that Lorentz-violating effects are, in a like manner, also small. Cohen and Glashow hasten to add that there are alternate ways of incorporating $CP$ violation without invoking Lorentz symmetries. This is done in the SM by the Cabibbo-Kobayashi-Maskawa mixing matrix mechanism. 

Cohen and Glashow suggest that VSR operates at the Planck scale and that the SM emerges as an effective theory from a more fundamental VSR theory. We propose that VSR is more likely to become operative not at higher energies, but at energies similar to those of the SM (but see remarks in section\;\ref{Sec:VSRremarks}). We are thus led to consider dark matter as the natural place for VSR. It is this argument that we explore here.
 
We briefly review VSR, and then construct a VSR invariant mass dimension one  fermionic quantum  field of spin one half. This field has an intrinsic darkness with respect to the standard model fields and has striking parallels to the previous work on \textbf{E}igenspinoren des \textbf{L}adungs\textbf{k}onjugations\textbf{o}perators (eigenspinors of the charge conjugation operator, Elko)~\cite{Ahluwalia:2004sz,Ahluwalia:2004ab,daRocha:2005ti,Ahluwalia:2008xi,Fabbri:2009ka,Boehmer:2010ma}\footnote{Both the Elko and the VSR theories have significant literature on their physical implications and their mathematical structure. To review them here will take us far afar from the task at hand. For this reason we provide an expanded list of references~\cite{Silagadze:2010bi,Shankaranarayanan:2009sz,Wei:2010ad,Boehmer:2007dh,Boehmer:2007ut,Kouretsis:2008ha,Vacaru:2010fi,Hariton:2006zj,Muck:2008bd,Das:2010cn,daRocha:2008we,Boehmer:2006qq,Alvarez:2008uy,Cohen:2006sc,SheikhJabbari:2008nc,Li:2010zb,Fabbri:2009aj,Cheon:2009zx,SheikhJabbari:2008wy,Cohen:2006ir,Das:2009fi,Bernardini:2007ex,Liberati:2009uq,Bernardini:2008ef,Gredat:2008qf,daRocha:2007pz} \textendash~with the general flavour of the works becoming obvious from their titles.}. We shall call the new construct Elko because its VSR variant carries the same defining features as the original construct and places the original formalism on a firmer theoretical footing.

\section{Review of VSR: the vector representation}
\label{Sec:VSR-Review}
 
VSR, by its defining features, is restricted to four subgroups of the Lorentz group. While each of these has quite a  different character, they all share the defining property that incorporating either $P$, $T$, $CP$, or $CT$ enlarges these subgroups to the full Lorentz group. With $T_1\colonequals K_x + J_y$,  $T_2\colonequals K_y - J_x$, where $\J$ and $\K$ are the generators of rotations and boosts, respectively, the algebras underlying these, up to isomorphisms, are enumerated in Table~\ref{table:1}.
\begin{table}[!hbt]
\centering
\begin{tabular}{lll}
\toprule
Designation & Generators & Algebra \\
\midrule
$\mathfrak{t}(2)$ & $T_1,T_2$ & $[T_1,T_2]=0$ \\
$\mathfrak{e}(2)$ & $T_1,T_2,J_z$ & $[T_1,T_2]=0,\  [T_1,J_z] = - i T_2,\ [T_2,J_z] = i T_1$ \\
$\mathfrak{hom}(2)$ & $T_1,T_2,K_z$ & $[T_1,T_2]= 0,\ [T_1,K_z] = i T_1,\  [T_2,K_z] = i T_2$ \\
$\mathfrak{sim}(2)$ & $T_1,T_2,J_z,K_z$ & $[T_1,T_2]= 0,\ [T_1,K_z] = i T_1,\  [T_2,K_z] = i T_2$ \\
& & $[T_1,J_z] = - i T_2,\  [T_2,J_z] = i T_1,\  [J_z,K_z] = 0$ \\ 
\bottomrule
\end{tabular}
\caption{The four VSR algebras.}
\label{table:1}
\end{table}

Apart from the discrete symmetries of \emph{C} and \emph{PT}, the VSR adjoins these subgroups with four spacetime translations, and introduces three new parameters associated with $T_1$, $T_2$, and $K_z$
\begin{align}
& \epsilon \ce \frac{p_x}{E-p_z} \\
& \varepsilon \ce \frac{p_y}{E-p_z} \\
& \varsigma \ce -\ln \left( \frac{E-p_z}{m}\right) \end{align}
respectively. In regard to the discrete symmetries \emph{P}, \emph{T}, and \emph{C}: any of the first two enlarges the four VSR avatars  to the full Lorentz group. Therefore, VSR only allows the incorporation of the charge conjugation symmetry. The \emph{HOM}(2) and \emph{SIM}(2) transformation 
\begin{equation}
\Lambda \ce e^{i T_1 \epsilon} e^{i T_2 \varepsilon} e^{i K_z \varsigma}
\label{eq:Lambda}
\end{equation}
takes the standard energy momentum four vector $k^\mu:=(m,0,0,0)$ into a general energy momentum four vector $p^\mu\colonequals(E,p_x,p_y,p_z)$. In the vector representation $T_1^3 =T_2^3=\mathbb{O}_4$ and $K_z^3 = -K_z$. This allows us to immediately expand the exponentials in eq.~(\ref{eq:Lambda})
\begin{align}
&  e^{i T_1 \epsilon} = \openone_4 + i T_1\epsilon - \frac{1}{2} T_1^2 \epsilon^2 \quad (\mbox{and a similar expression for}\ e^{i T_2 \varepsilon})\\
& e^{i K_z \varsigma} = \openone_4 + i K_z \sinh\varsigma + K_z^2 (1-\cosh\varsigma) 
\end{align}
The symbols $\openone_n$ and $\mathbb{O}_n$ represent the $n\times n$ identity and the null matrices, respectively. With these expansions at hand eq.~(\ref{eq:Lambda}) becomes
\beq
\Lambda = \begin{pmatrix}
\frac{E}{m} & \frac{p_x}{E-pz} & \frac{p_y}{E-p_z} & \frac{m^2-E(E-p_z)}{m(E-p_z)} \\
\frac{p_x}{m} & 1 & 0 & -\frac{p_x}{m} \\
\frac{p_y}{m} & 0 & 1 & -\frac{p_y}{m} \\
\frac{p_z}{m} & \frac{p_x}{E-pz} & \frac{p_y}{E-p_z} & \frac{m^2-p_z(E-p_z)}{m(E-p_z)} \\
\end{pmatrix}\label{Eq:vsrlambdaVector}
\eeq
On setting $E=\gamma m$, $p_i = \gamma m u_i$ ($i\in\{1,2,3\}$), the above result coincides with its counterpart obtained by Das and Mohanty~\cite[eq. 9]{Das:2009fi}. For the sake of completeness and later use, we also note that because $J_z^3 = J_z$ (in the vector representation)
\beq
e^{i J_z \varphi}  = \openone_4 + i J_z \sin\varphi + J_z^2(\cos\varphi-1)
\eeq
One can now transform two events $x_1^\mu\ce (t_1,\x)$ and $x_2^\mu\ce (t_2,\x)$ by the $\Lambda$ transformation so obtained, and immediately verify that 
\beq
t_2^\prime - t_1^\prime = \gamma_\mathbf{u} (t_2 -t_1),
\eeq
where $\gamma_\mathbf{u}\ce 1/\sqrt{1-\mathbf{u}^2}$. In the process one confirms that indeed VSR reproduces the SR result on time dilation. Similarly, invariance of the interval, and thus the existence of a universal and isotropic velocity, holds for all observers connected by a \emph{HOM}(2) or \emph{SIM}(2) VSR transformation. Therefore, following Cohen and Glashow we remind that invariance under \emph{HOM}(2) is both necessary and sufficient to ensure that there is an invariant  and universal speed  for all observers. 

\section{Spinor representations of $\mathfrak{hom}(2)$ and $\mathfrak{sim}(2)$ algebra underlying VSR} \label{Sec:spinor-representation} 

To distinguish the spinor representation from the above-considered vector representation, we adapt the notation as follows
\beq
T_1\to \tau_1,\ T_2\to\tau_2,\ K_z\to\kappa_z,\ J_z\to\zeta_z \quad(\mbox{and} \;\K\to\k,\ \J\to\z)
\eeq
The $\mathfrak{hom}(2)$ spinor representations are very similar to those of $\mathfrak{sim}(2)$ except that they do not have to respect the rotational symmetry under the preferred direction. Their construction is not given explicitly in what follows. 

For spin one half the VSR $\mathfrak{sim}(2)$ algebra admits two types of spinor representations with the respective generators given by 
\begin{itemize}
\item Type a:  $\tau_1^a\ce \kappa_x^a+\zeta_y^a$, $\tau_2^a\ce \kappa_y^a-\zeta_x^a$,  $\zeta_z^a$, and $\kappa_z^a$\\
 where  $\k^a = -i  \s/2$, and $\z^a = \s/2$.
\item Type b:  $\tau_1^b\ce \kappa_x^b+\zeta_y^b$, $\tau_2^b\ce \kappa_y^b-\zeta_x^b$, $\zeta_z^b$, and $\kappa_z^b$\\
where  $\k^b = + i  \s/2$, and $\z^b = \s/2$.
\end{itemize}
As will be shown below, Elko reside in a representation space that is the direct sum of the a-type and the b-type spaces. The VSR generators for this four dimensional spinor representation space are 
\beq
\tau_1\ce \tau^a_1\oplus\tau^b_1,\ \tau_2\ce \tau^a_2\oplus\tau^b_2,\ \zeta_z\ce\zeta_z^a\oplus\zeta_z^b,\ \kappa_z\ce\kappa_z^a\oplus\kappa_z^b
\eeq
The \emph{SIM}(2) transformation that takes the standard four-component VSR spinor $\chi(k^\mu)$ into a general VSR spinor $\chi(p^\mu)$ is
\begin{subequations}
\beq \lambda \ce 
e^{i \tau_1 \epsilon}
e^{i \tau_2 \varepsilon} 
e^{i \kappa_z \varsigma} 
\label{eq:lambda}
\eeq
For sake of later reference we define
\beq
\Gamma_1(\epsilon)\ce e^{i \tau_1 \epsilon},\qquad
\Gamma_2(\varepsilon)\ce e^{i \tau_2 \varepsilon},\qquad
\Gamma_3(\varsigma)\ce e^{i \kappa_z\varsigma}
\eeq
\end{subequations}
 In the spinor representation $\tau_1^2 =\tau_2^2=\mathbb{O}_4$, and  $(2 \kappa_z)^3 = - 2 \kappa_z$. This allows us to immediately expand the exponentials in eq.~(\ref{eq:lambda})
\begin{align}
& e^{i \tau_1\epsilon} = \openone_4 + i \tau_1\epsilon \quad (\mbox{with a similar expression for}\ e^{i \tau_2 \varepsilon}) \\
& e^{i \kappa_z \varsigma} = \openone_4 + i  2 \kappa_z \sinh(\varsigma/2) + 4( \kappa_z)^2 (1-\cosh(\varsigma/2)) 
\end{align}
After not too lengthy a calculation, these results yield
\beq
\lambda=\begin{pmatrix} 
\sqrt{\frac{m}{E-p_z} } & \frac{p_x-i p_y}{\sqrt{m (E-p_z)}} & 0 & 0 \\
0 & \sqrt{\frac{E-p_z}{m} } & 0 & 0 \\
0 & 0 & \sqrt{\frac{E-p_z}{m} } & 0 \\
0 & 0 &   -\frac{p_x+i p_y}{\sqrt{m (E-p_z)}} & \sqrt{\frac{m}{E-p_z}} \\
\end{pmatrix}\label{eq:lambdafull}
\eeq
We note that $(2 \zeta_z)^3 = 2 \zeta_z$. As a result
\begin{subequations}
\beq
e^{i \zeta_z \varphi} = \openone_4 + i 2 \zeta_z \sin(\varphi/2) + 4 \zeta_z^2(\cos(\varphi/2)-1)
\eeq
For later reference we define
\beq
\Gamma_4(\varphi)\ce e^{i \zeta_z \varphi}
\eeq
\end{subequations}

\subsection{VSR-spinor dual} 
\label{Sec:Dual}

To introduce spinor fields in VSR we will need an adjoint. Towards that task we now introduce a dual to $\chi(\p^\mu)$
\beq
\dual{\chi}(p^\mu)\ce \left[\chi(\p^\mu)\right]^\dagger \gamma
\eeq
with $\gamma$ determined by the condition that  $\dual{\chi}(p^\mu)\,\chi(p^\mu)$ is invariant under any of the \emph{SIM}(2) VSR transformations. This gives the following commutator and anticommutator conditions on $\gamma$
\begin{subequations}
\begin{align}
& [\zeta_z,\gamma]=0,\ \{\kappa_z,\gamma\}=0,\ [\tau_1,\gamma] = 2 \kappa_x \gamma,\\
& [\tau_2,\gamma] = 2 \kappa_y \gamma,\ 
\{\tau_1 \tau_2 \kappa_z, \gamma\} = (\kappa_z \kappa_y \zeta_y - \kappa_z\zeta_x \kappa_x) 2 \gamma
\end{align}
\end{subequations}
The unique solution to these constraints is 
\beq
\gamma=\begin{pmatrix}
0 & 0 & \alpha & 0 \\
0 & 0 & 0 &\alpha \\
\beta & 0 & 0 & 0 \\
0 & \beta & 0 & 0
\end{pmatrix} 
\eeq
with $\alpha,\beta \in \mathbb{C}$. To treat a- and b- type spinors on the same footing, and to keep the norm real, we set $\alpha=\beta=\pm \,i$ \footnote{It is also to be noted that while $\gamma$ formally appears to be the same as the $\eta$ of ref.~\cite[cf. eq. (16)]{Ahluwalia:2008xi} the constraints it satisfies are quite different.}. Thus there are two options for $\gamma$
\beq
\gamma_\pm  \ce \pm\, i\,\gamma^0
\eeq

\subsection{Standard spinors in VSR} \label{Sec:standard-spinors}

At this exploratory stage we are guided by the fact that Dirac spinors are eigenspinors of the parity operator. However, the VSR framework does not allow any of the $P$, $T$, $CP$ or $CT$ symmetries. Their incorporation, as already emphasised, immediately requires an enlargement of the VSR subgroups to the Lorentz group. For this reason we turn to the charge conjugation operator~\cite{Ahluwalia:2004ab}
\beq
C \ce \begin{pmatrix}
0 & i \Theta \\
-i\Theta & 0
\end{pmatrix}\mathcal{K}\label{eq:chconj}
\eeq
where $\mathcal{K}$ is the complex conjugation operator (while acting to the right), and $\Theta$ is the spin one half Wigner time reversal operator defined by $\Theta (\s/2) \Theta^{-1} = - (\s/2)^\ast$. We use the representation
\beq
\Theta = \begin{pmatrix}
0 & -1 \\
1 & 0
\end{pmatrix}
\eeq
To construct the standard VSR spinor, we make the observation that if $\Gamma^a$ and $\Gamma^b$ represent any of the \emph{SIM}(2) VSR transformations on the a-type and the b-type spinors, respectively, then
\beq
\left(\Gamma^b\right)^\ast = \Theta^{-1} \Gamma^a \Theta
\eeq
In addition,  $C$ and $\Gamma\ce\Gamma^a\oplus\Gamma^b$ commute
\begin{equation}
[C,\Gamma]=0\label{eq:cxi}
\end{equation}
\emph{Equipped with these observations it can be easily shown that if $\chi^b(k^\mu)$ is a b-type bispinor, then $\wp \,\Theta \left[\chi^b(k^\mu)\right]^\ast$ transforms as an a-type bispinor (here, $\wp:=e^{i \delta},\ \delta\in\mathbb{R}$).} Thus there exists a natural standard four-component \emph{SIM}(2) VSR spinor 
\beq
\chi(k^\mu)\ce \begin{pmatrix}
\wp\, \Theta\left[\chi^b(k^\mu)\right]^\ast \\ 
\chi^b(k^\mu)
\end{pmatrix}
\eeq
The unknown phase factor $\wp$ is now fixed by the condition that $\chi(k^\mu)$ be an Elko, i.e., an eigenspinor of the charge conjugation operator with real eigenvalues: $C\chi(k^\mu) = \pm \chi(k^\mu)$. This yields, $\wp= \pm\, i$. We shall denote the self conjugate spinors by the symbol $\rho(k^\mu)$, and the anti-self conjugate spinors by $\varrho(k^\mu)$. Setting $\Gamma=\lambda$ in eq.~(\ref{eq:cxi}) we see that the charge conjugation operator commutes with the transformation $\lambda$, which takes $k^\mu$ to $p^\mu$ (we parameterise $p^\mu$ as $(E, p \sin\theta\cos\phi, p \sin\theta\sin\phi,p\cos\theta)$). For this reason not only the standard $\chi(k^\mu)$ but also the $\chi(p^\mu) \ce \lambda\, \chi(k^\mu)$ is an Elko. The space in which $\chi^b(k^\mu)$ resides is two dimensional. We span this space by eigenspinors of the $\zeta_z^b$~\footnote{All that has been said up to this point applies to \emph{HOM}(2) as well as to \emph{SIM}(2). However, $\zeta_z^b$  is not a generator of \emph{HOM}(2), so that what follows is specific to \emph{SIM}(2).}. These read
\beq
\chi_+^b(k^\mu)\ce e^{i \vartheta_+} \sqrt{m} \begin{pmatrix}1 \\ 0\end{pmatrix} ,\quad
\chi_-^b(k^\mu)\ce  e^{i \vartheta_-}\sqrt{m} \begin{pmatrix}0 \\ 1\end{pmatrix} 
\label{eq:chipm}
\eeq
We fix the $\vartheta_\pm$ phases by demanding that the quantum fields, to be discussed below, carry minimal departures from locality (see section\;\ref{Sec:locality}); this gives $\vartheta_\pm = \mp \phi/2$ \footnote{The fact that we do not carry these phases in what follows and fully show their propagation, and their constraining, is to avoid presenting certain details that are suitable for research notes but unenlightening and cumbersome for a research communication.}. As we will show, this choice of phases will yield a theory that is invariant under \emph{SIM}(2) VSR, while it has been shown in the past that it is not Lorentz invariant~\cite{Ahluwalia:2008xi,Boehmer:2010ma}. Thus we obtain four \emph{SIM}(2) VSR spinors
\begin{equation}
\chi(k^\mu) \to \left\{
\begin{array}{l}
\rho_\pm(k^\mu) \ce \rho(k^\mu) \vert_{\chi^b\to\chi^b_\pm} \\
\varrho_\pm(k^\mu) \ce \pm \varrho(k^\mu) \vert_{\chi^b\to\chi^b_\mp} \\
\end{array} \right.
\end{equation}
With $\lambda$ given by eq.~(\ref{eq:lambdafull})
\beq
\chi(p^\mu) = \lambda \,\chi(k^\mu)\eeq 
While the $\rho(k^\mu)$ and $\varrho(k^\mu)$ are identical to the $\xi(\0)$ and $\zeta(\0)$ of ref.~\cite{Ahluwalia:2008xi}, respectively, the $\rho(p^\mu)$ and $\varrho(p^\mu)$ are not identical to the $\xi(\p)$ and $\zeta(\p)$. This is because the spinor boost associated with the Lorentz group~\textendash~denoted by $\kappa$ in  ref.~\cite[eq. 14]{Ahluwalia:2008xi}~\textendash~is not the same as its counterpart $\lambda$ of the \emph{SIM}(2) VSR.

The results of section\;\ref{Sec:Dual} now allow us to introduce the dual to the \emph{SIM}(2) VSR spinors
\begin{align}
\dual\rho_\pm(p^\mu) &\ce - \left[\rho_\mp(p^\mu)\right]^\dagger\,\gamma_\pm\\
\dual\varrho_\pm(p^\mu) &\ce - \left[\varrho_\mp(p^\mu)\right]^\dagger\,\gamma_\pm
\end{align}

\subsection{\emph{SIM}(2) VSR spinors do not satisfy Dirac equation}
\label{Sec:Dirac}

We now examine the action of the Dirac operator $\dirac$ on the \emph{SIM}(2) VSR spinors. Unlike the Dirac spinors of SR the \emph{SIM}(2) VSR spinors are not the eigenspinors of the $\dirac$ operator:\footnote{As a parenthetic remark we note that the VSR invariant equation for neutrinos given in ref.~\cite{Cohen:2006ir} suffers from the fact that $\mathrm{det} \left(\dirac- \frac{m^{2}}{2}\frac{{\gamma_\mu n^\mu}}{p_\mu n^\mu}\right)=0 $, where $n^\mu\colonequals(1,0,0,1)$, leads to a non-Einsteinian  dispersion relation. We do not pursue this problem here. We also note that in the context of original Elko a result similar to the one contained in eq.~(\ref{eq:dvoeglazov}) was obtained in ref.~\cite{Dvoeglazov:1995kn}.}
\beq
\dirac \rho_\pm (p^\mu) \not\propto \rho_\pm (p^\mu)\label{eq:dvoeglazov}
\eeq 
Instead, a direct evaluation reveals: 
\begin{subequations}
\begin{align}
\dirac \rho_\pm(p^\mu) = \pm\, i m \rho_\mp(p^\mu)\\
\dirac \varrho_\pm(p^\mu) = \mp\, i m \varrho_\mp(p^\mu)
\end{align}
\end{subequations}
Operating on $\dirac \rho_+(p^\mu) = + \, i m \rho_-(p^\mu)$ with $\gamma_\nu p^\nu$ from the left, using $\gamma_\nu p^\nu \rho_-(p^\mu) = - \, i m \rho_+(p^\mu)$, and recalling that $\{\gamma^\mu,\gamma^\nu\}= 2 \eta^{\mu\nu}$, yields the spinor Klein-Gordon equation $\left[\left(p_\mu p^\mu - m^2 \right)\otimes\openone_4\right]\rho_+(p^\mu) = 0.$ Repeating the same excercise for the other spinors confirms that all the \emph{SIM}(2) VSR spinors satisfy the momentum-space Elko Klein-Gordon equation $\left[\left(p_\mu p^\mu - m^2 \right)\otimes\openone_4\right]\rho_\pm(p^\mu) = 0$ and $\left[\left(p_\mu p^\mu - m^2 \right)\otimes\openone_4\right]\varrho_\pm(p^\mu) $ $ = 0$. Since the spacetime translations are still a symmetry of VSR we can use $p^\mu \to i\partial^\mu$ in the context of \emph{SIM}(2) VSR. As a consequence, Elko in \emph{SIM}(2) VSR  satisfy the spinor Klein-Gordon equation
\beq
\left[\left( \eta_{\mu\nu}\partial^\mu\partial^\nu + m^2 \right)\otimes \openone_4\right] \rho_\pm(x) =0,\quad
\left[\left(\eta_{\mu\nu}\partial^\mu\partial^\nu + m^2 \right)\otimes\openone_4\right]\varrho_\pm(x) =0
\eeq
where $\rho_\pm(x)\ce \rho_\pm(p^\mu) e^{-i p^\mu x_\mu}$ and $\varrho_\pm(x)\ce \varrho_\pm(p^\mu) e^{+i p^\mu x_\mu}$. The above equations are \emph{SIM}(2) VSR invariant, but not Lorentz invariant. The reason for this is that while the spinor Klein-Gordon operator is Lorentz invariant, the VSR Elko are only \emph{SIM}(2) VSR invariant.

\section{ \emph{SIM}(2) VSR invariant spin-half fermionic fields and their natural darkness} 
\label{Sec:qf}

With the calculations presented above, and with the insights gained from our previous work on Elko~\cite{Ahluwalia:2004sz,Ahluwalia:2004ab,Ahluwalia:2008xi}, we introduce two new  spin-half fermionic fields (justification for various factors in the integration measure are similar to those given in ref.~\cite[section 7]{Ahluwalia:2004ab}):
\beq
\Upsilon(x)\ce\int\frac{\ud^3 p}{(2\pi)^3} \frac{1}{\sqrt{2 m E(\p)}}\sum_{\alpha=+,-}\left[a_\alpha(\p) 
 \rho_\alpha(\p)  e^{-i p^\mu x_\mu} + b^\dag_\alpha(\p) 
 \varrho_\alpha(\p)  e^{+i p^\mu x_\mu}  \right]\label{eq:Upsilonb}
\eeq
and its neutral counterpart
\beq
\upsilon(x)\ce\int\frac{\ud^3 p}{(2\pi)^3} \frac{1}{\sqrt{2 m E(\p)}}\sum_{\alpha=+,-}\left[a_\alpha(\p) 
 \rho_\alpha(\p)  e^{-i p^\mu x_\mu} + a^\dag_\alpha(\p) 
 \varrho_\alpha(\p)  e^{+i p^\mu x_\mu}  \right]\label{eq:upsilona}
\eeq
For reasons which parallel those given in ref.~\cite[section 7]{Ahluwalia:2004ab} we have the following anticommutators
\begin{subequations}
\begin{align}
&\{ a_\alpha(\p), a_{\alpha'}^\dag(\p') \} = (2 \pi)^3 \delta^3(\p - \p') \delta_{\alpha \alpha'} \\
&\{ a_\alpha(\p), a_{\alpha'}(\p') \} = 0,\qquad \{ a_\alpha^\dag(\p), a_{\alpha'}^\dag(\p') \} = 0
\end{align}
\end{subequations}
with similar anticommutators for $b_\alpha(\p)$ and  $b^\dagger_\alpha(\p)$. The adjoint fields are defined as
\begin{align}
&\dual{\Upsilon}(x)\ce\int\frac{\ud^3 p}{(2\pi)^3} \frac{1}{\sqrt{2 m E(\p)}}\sum_{\alpha=+,-}\left[a^\dagger_\alpha(\p) \dual{\rho}_\alpha(\p) e^{+i p^\mu x_\mu} + b_\alpha(\p) \dual{\varrho}_\alpha(\p) e^{-i p^\mu x_\mu} \right] \\
&\dual{\upsilon}(x)\ce\dual{\Upsilon}(x)\big\vert_{b_\alpha(\mathbf{p})\to a_\alpha(\mathbf{p})}
\end{align}
The discussion of section\;\ref{Sec:Dirac} has the consequence that it is the spinor Klein-Gordon operator, rather than the Dirac operator, that annihilates $\Upsilon(x)$ and $\upsilon(x)$. The associated Lagrangian densities 
\begin{subequations}
\begin{align}
&{\mathcal{L}}^\Upsilon(x)= \partial^\mu\dual{\Upsilon}(x)\partial_\mu\Upsilon(x) - m^2 \dual{\Upsilon}(x) \Upsilon(x)\label{eq:LUpsilon} \\
&{\mathcal{L}}^\upsilon(x)= \partial^\mu\dual{\upsilon}(x)\partial_\mu\upsilon(x) - m^2 \dual{\upsilon}(x) \upsilon(x)\label{eq:Lupsilon}
\end{align}
\end{subequations}
confer a mass dimension of one for both of these fermionic fields of spin one half. This is a rather remarkable result and places the new fields on an entirely new footing as it restricts the type of gauge fields they can support and, consequently, limits the type of allowable interactions between these new fields and the fields of the SM\footnote{Further justification for (\ref{eq:LUpsilon}) and (\ref{eq:Lupsilon}) can be derived by directly evaluating the time-ordered products, $\langle~\vert \mathcal{T_\#} [\Upsilon(x^\prime) \dual{\Upsilon}(x)]\vert~\rangle$ and $\langle~\vert \mathcal{T_\#} [\upsilon(x^\prime) \dual{\upsilon}(x)]\vert~\rangle$ respectively. Here $\mathcal{T_\#}$ is the Elko time-ordering operator introduced in ref.~\cite[Appendix A]{Ahluwalia:2009rh}.}. For instance, the quartic self interaction of the mass dimension three half fermionic fields of the SM are suppressed by two powers of the unification scale\;\textendash\;usually taken as the Planck scale. Similar self interactions for the new fields are completely unsuppressed  as they are described by dimension four operators. The point-interaction Lagrangian densities that couple the fermionic fields of the SM with the new fields are dimension five operators; thereupon, interactions between the new fields and the fields of the SM are again suppressed.

Therefore, there is an urgent need to develop new calculational tools to investigate signatures of the new VSR Elko fields at the Large Hadron Collider. These are required by the new locality structure to be discussed below in section\;\ref{Sec:locality}.
 
\subsection{Orthormality relations, spin sums, twisted spin sums, and completeness for \emph{SIM}(2) VSR spinors}

In order to study the locality structure of the introduced fields we need various identities. These follow under appropriate headings.

\paragraph{Orthonormality relations}
\begin{subequations}
\begin{align}
&\dual{\rho}_\alpha(\p) \rho_{\alpha^\prime}(\p) = + 2 m\ \delta_{\alpha\alpha^\prime} \\
&\dual{\varrho}_\alpha(\p) \varrho_{\alpha^\prime}(\p) = - 2 m\ \delta_{\alpha\alpha^\prime} \\
&\dual{\rho}_\alpha(\p) \varrho_{\alpha^\prime}(\p) = \dual{\varrho}_\alpha(\p) \rho_{\alpha^\prime}(\p) = 0
\end{align}
\end{subequations}

\paragraph{Spin sums}
\begin{subequations}
\begin{align}
&\sum_\alpha \rho_\alpha(\p) \dual{\rho}_\alpha(\p) = m \left[ \mathcal{G}(\p) +\openone_4\right] \label{eq:ssrho} \\
&\sum_\alpha \varrho_\alpha(\p) \dual{\varrho}_\alpha(\p) = m \left[ \mathcal{G}(\p) -\openone_4\right]\label{eq:ssvarrho}
\end{align}
\end{subequations}
where
\beq
\mathcal{G}(\p)\ce i
\begin{pmatrix}
0 & 0 & 0 &-e^{-i\phi} \\
0 & 0 & e^{i\phi} & 0 \\
0 & -e^{-i\phi} & 0 & 0 \\
e^{i\phi} & 0 & 0 & 0
\end{pmatrix}
\eeq
is an odd function of $\p$: $\mathcal{G}(\p) = - \mathcal{G}(-\p)$. Equivalently, given that $\mathcal{G}(\p)$ depends only on $\phi$ and not on $\theta$ and $p$, 
\beq
\mathcal{G}(\phi) = - \mathcal{G}(\pi+\phi)\label{eq:gphi}
\eeq

In a Lorentz invariant theory the counterpart of $\mathcal{G(\p)}$, as our reader surely knows, is $m^{-1} \gamma^\mu p_\mu$. It can be arrived at in precisely the same manner as $\mathcal{G(\p)}$ here. To examine invariance of our theory under VSR, we note by direct evaluation that
\begin{subequations}
\begin{align}
\Gamma_1(\epsilon) \mathcal{G}(\phi)\left[\Gamma_1(\epsilon)\right]^{-1} \Big\vert_{\varepsilon=0,\varsigma=0,\varphi=0}
&= \mathcal{G}(\phi)  \label{eq:vsr1} \\
\Gamma_2(\varepsilon) \mathcal{G}(\phi)\left[\Gamma_2(\varepsilon)\right]^{-1} \Big\vert_{\epsilon=0,\varsigma=0,\varphi=0}
&= \mathcal{G}(\phi)  \label{eq:vsr2} \\
\Gamma_3(\varsigma) \mathcal{G}(\phi)\left[\Gamma_3(\varsigma)\right]^{-1} \Big\vert_{\epsilon=0,\varepsilon=0,\varphi=0}
&= \mathcal{G}(\phi)  \label{eq:vsr3} 
\end{align}
and that
\beq
\Gamma_4(\varphi) \mathcal{G}(\phi)\left[\Gamma_4(\varphi)\right]^{-1} \Big\vert_{\epsilon=0,\varepsilon=0,\varsigma=0}
= \mathcal{G}(\phi+\varphi)  \label{eq:vsr4} 
\eeq
\end{subequations}
As a result $\mathcal{G}(\phi)$ is invaraint under the \emph{HOM}(2) transformations, and covariant (i.e., form invariant) under the \emph{SIM}(2) transformations.

\paragraph{Twisted spin sums}
\begin{subequations}
\begin{align}
\sum_\alpha \Big[\rho_\alpha(\p) {\varrho}^\mathrm{T}_\alpha(\p)+ \varrho_\alpha(- \p) {\rho}^\mathrm{T}_\alpha(-\p)\Big] &= \mathbb{O}_4 
\label{eq:ss49} \\
\sum_\alpha \left[ \left(\dual{\rho}_\alpha(\p) \right)^\mathrm{T} \dual{\varrho}_\alpha(\p) + (\dual{\varrho}_\alpha(- \p))^\mathrm{T} \dual{\rho}_\alpha(-\p) \right] &= \mathbb{O}_4\label{eq:after50}
\end{align}
\end{subequations}
where the  superscript  $\mathrm{T}$ denotes the matrix transpose.

\paragraph{Completeness relation}
\beq
\frac{1}{2m}\sum_\alpha\Big[ \rho_\alpha(\p) \dual{\rho}_\alpha(\p) - \varrho_\alpha(\p) \dual{\varrho}_\alpha(\p) \Big] =\openone_4
\label{eq:compl}
\eeq

\subsection{Locality structure of VSR Elko fields}
\label{Sec:locality}

With the required identities evaluated, we now return to present the locality structure of the VSR Elko fields introduced in section\;\ref{Sec:qf}. The canonically conjugate momenta to $\Upsilon(x)$ and $\upsilon(x)$ are
\beq
\Pi(x) \ce \frac{\partial{\mathcal{L}^\Upsilon}}{\partial\dot\Upsilon} = \frac{\partial}{\partial t} \dual{\Upsilon}(x),\qquad
\pi(x) \ce \frac{\partial{\mathcal{L}^\upsilon}}{\partial\dot\upsilon} = \frac{\partial}{\partial t} \dual{\upsilon}(x)
\eeq
In what follows we treat each of the introduced fields in turn.

\subsubsection{Locality structure of  $\Upsilon(x)$}

Using the above results and definitions, we calculate the equal-time anticommutator of $\Upsilon(x)$ with $\Pi(x)$. This yields
\beq
\left\{\Upsilon(\x,t),\ \Pi(\x^\prime,t) \right\} = i\int \frac{\ud^3 p}{(2\pi)^3}\frac{1}{2 m} e^{i\mathbf{p}\cdot(\mathbf{x}-\mathbf{x^\prime})} \sum_\alpha\Big[ \rho_\alpha(\p) \dual{\rho}_\alpha(\p) - \varrho_\alpha(-\p) \dual{\varrho}_\alpha(-\p) \Big] 
\eeq
Using eq.~(\ref{eq:ssrho}) and eq.~(\ref{eq:ssvarrho}), and that $\mathcal{G}(\p)$ is an odd function of $\p$, we readily find
\beq
\sum_\alpha\Big[ \rho_\alpha(\p) \dual{\rho}_\alpha(\p) - \varrho_\alpha(-\p) \dual{\varrho}_\alpha(-\p) \Big] 
= 2 m \left[ \openone_4 + \mathcal{G}(\p)\right]
\eeq
Thus
\beq
\left\{\Upsilon(\x,t),\ \Pi(\x^\prime,t) \right\} = i\delta^3(\x-\x^\prime)\openone_4 + \underbrace{i\int \frac{\ud^3 p}{(2\pi)^3} \ \mathcal{G}(\p) e^{i\mathbf{p}\cdot(\mathbf{x}-\mathbf{x^\prime})}}_{\equalscolon I_{\mathcal{G}}}
\eeq
This is adjoined by the field-field and momentum-momentum anticommutators
\beq
\left\{\Upsilon(\x,t),\ \Upsilon(\x^\prime,t) \right\} =\mathbb{O}_4,\qquad
\left\{\Pi(\x,t),\ \Pi(\x^\prime,t) \right\} =\mathbb{O}_4
\eeq
$I_\mathcal{G}$ identically vanishes for $\x-\x^\prime$ parallel to the $z$-axis. In this case  the inner product of $\x-\x^\prime$ with $\p$ becomes independent of $\phi$, while integration of $\mathcal{G}(\phi)$ \textendash~see, eq.~(\ref{eq:gphi})~\textendash~over one period vanishes. As a result, along the VSR preferred axis, chosen as $z$ here, $\Upsilon(x)$ is local. This minimal departure from locality is therefore entirely encoded in $I_\mathcal{G}$.

\subsubsection{Locality structure of $\upsilon(x)$}

Following as above, we find that
\beq
\left\{\upsilon(\x,t),\ \pi(\x^\prime,t) \right\} = i\delta^3(\x-\x^\prime)\openone_4 + I_\mathcal{G} \label{eq:f-m-upsilon}
\eeq
while the field-field anticommutator takes the form
\beq
\left\{\upsilon(\x,t),\ \upsilon(\x^\prime,t) \right\} = i\int \frac{\ud^3 p}{(2\pi)^3}\frac{1}{2 m E(\mathbf{p})} e^{i\mathbf{p}\cdot(\mathbf{x}-\mathbf{x^\prime})} \sum_\alpha \Big[\rho_\alpha(\p) {\varrho}^\mathrm{T}_\alpha(\p)+ \varrho_\alpha(- \p) {\rho}^\mathrm{T}_\alpha(-\p)\Big]\nonumber
\eeq
Using eq.~(\ref{eq:ss49}) it reduces to 
\beq
\left\{\upsilon(\x,t),\ \upsilon(\x^\prime,t) \right\} =\mathbb{O}_4\label{eq:upsilonab}
\eeq
Similarly, the momentum-momentum anticommutator evaluates to 
\begin{align}
\left\{\pi(\x,t),\ \pi(\x^\prime,t) \right\} &= i\int \frac{\ud^3 p}{(2\pi)^3}\frac{E(\mathbf{p})}{2 m } e^{- i\mathbf{p}\cdot(\mathbf{x}-\mathbf{x^\prime})}\nonumber \\
&\quad \times \sum_\alpha \left[ \left(\dual{\rho}_\alpha(\p) \right)^\mathrm{T} \dual{\varrho}_\alpha(\p) + (\dual{\varrho}_\alpha(- \p))^\mathrm{T} \dual{\rho}_\alpha(-\p) \right] 
\end{align}
Using eq.~(\ref{eq:after50}) it reduces to 
\beq
\left\{\pi(\x,t),\ \pi(\x^\prime,t) \right\} =\mathbb{O}_4 \label{eq:pi}
\eeq
Like its sibling $\Upsilon(x)$, $\upsilon(x)$ is also local along the VSR preferred direction. Moreover, the locality condition is again entirely dependent on $I_\mathcal{G}$ as is clearly evident from equations~(\ref{eq:f-m-upsilon}), (\ref{eq:upsilonab}), and (\ref{eq:pi}).

It is worth mentioning that the minimal departure from locality has been achieved by the judicious choice of $\vartheta_\pm$-dependent phases for the VSR standard spinors \textendash~see eq.~(\ref{eq:chipm}).

 \section{Before we conclude}
 
We now address some of the natural questions that arise in a systematic formulation of quantum field theories carrying VSR symmetries. These were deemed too distracting while the main formalism was developed above. The tone of the narrative now changes to suit the content and we do not hesitate to take a few brief detours.

\subsection{Elko and Majorana fermions: similarities and differences}
 
In 1937 Majorana introduced the idea that a new type of neutral field may be 
constructed from the Dirac field
\beq
\psi(x)\ce\int\frac{\ud^3 p}{(2\pi)^3} \frac{m}{p^0}\sum_{\sigma}\left[c_\sigma(\p) 
 u_\sigma(\p)  e^{-i p^\mu x_\mu} + d^\dag_\sigma(\p) 
 v_\sigma(\p)  e^{+i p^\mu x_\mu}  \right]
\eeq
by identifying the antiparticle creation operator $d^\dag_\sigma(\p) $ with the particle creation operator $c^\dag_\sigma(\p) $~\cite{Majorana:1937vz}
\beq
 \psi^M(x) \ce \psi(x)\big\vert_{d^\dag_\sigma({\mathbf p}) \rightarrow c^\dag_\sigma({\mathbf p})}
\eeq
In the Majorana field, the expansion coefficients of the field are the usual Dirac spinors. 
 
What is usually called a Majorana fermion in the high energy community, particularly in the literature on supersymmetry, is something different (though not entirely unrelated). In Weinberg's notation, this Majorana fermion has the form~\cite{Weinberg:2000cr}
\begin{equation}
  s \ce \left(
  \begin{array}{c}
    e\, \zeta^\ast \\ 
    \zeta
  \end{array}
  \right)
\end{equation}
where $e= i\sigma_2$. The fermionic aspect is introduced by demanding that $\zeta$ transforms as a spin one half spinor of the $\ell$-type, and that it be treated as a Grassmann number. Often it is said that such an object is a Weyl fermion in a four component form. This is because $\zeta$ has two independent degrees of freedom and as a result one only obtains two, and not four, four-component $s$'s.

Elko does not treat $\zeta$ as a Grassmann number, but, in the context of the Lorentz group, as an $\ell$-type complex-valued spinor, and, in the context of \emph{SIM}(2), as a b-type complex valued spinor. The fermionic aspect is introduced via the canonical operator formalism of quantum field theory. In the Elko formalism $e = \wp\, \Theta$, rather than $\pm i\sigma_2$, where $\wp$ is determined by the demand that $s$, now treated as a complex-valued spinor, be an eigenspinor of the charge conjugation operator (see, eq.~(\ref{eq:chconj})) with eigenvalues $\pm 1$. This places the Elko spinors at the same formal footing as the Dirac spinors. With these definitions and observations
\begin{enumerate}
  \item[\textemdash~] Elko manifestly violates rotational symmetry along the lines contained in VSR.
  \item[\textemdash~] Elko is not a Weyl spinor in disguise: under charge conjugation, there are now four independent spinors, two self conjugate, and, the other two, anti-self conjugate.
  \item[\textemdash~] The introduction of $\Theta$, the Wigner time reversal operator, allows an extension of the Elko concept to all fermionic fields of arbitrary half integral spin.
  \item[\textemdash~] For spin one half, $ \wp\, \Theta = \pm i \sigma_2=\pm e$.
  \item[\textemdash~] When Elko spinors are used to define \emph{SIM}(2) VSR invariant quantum fields as given in eq.~(\ref{eq:Upsilonb}) and eq.~(\ref{eq:upsilona}), Elko fields allow both the possibilities that  particles and antiparticles are distinct, as well as the possibility that particles and antiparticles are identical.
\end{enumerate}
The place of Elko within the Lounesto spinor classification is detailed in ref.~\cite{daRocha:2005ti}. A hint that Elko may require a non-commutative momentum space was noted in ref.~\cite[section 12.3 and Appendix B.6]{Ahluwalia:2004ab}.
 
\subsection{VSR, Elko, and non-commutative spacetime}
\label{Sec:VSRremarks}

Late in 1993~\cite{Ahluwalia:1993dd} it was realised that a merger of the Heisenberg fundamental commutator, and the gravitational effects in a measurement process, demands spacetime to be non-commutative (the argument was soon refined in~\cite{Doplicher:1994zv}). These effects become significant only at the Planck scale. Though motivated by different arguments, Sheikh-Jabbari and Tureanu find that the most natural realisation of (\emph{T}(2)) VSR happens in the setting of (Moyal) non-commutative spacetime~\cite{SheikhJabbari:2008nc}. Exploiting the notion of the Drinfel'd twist~\cite{Drinfeld:1983ky}, they show that inspite of the lack of full Lorentz symmetry, the fields carry representations of the full Lorentz group~\cite{Tureanu:2007zz,Chaichian:2007tk,Chaichian:2008ge} and the spin-statistics theorem is still valid. The deformation appears in the product of the fields, i.e. in the interaction terms. In a clean sweep, they thus provided a theoretically rigorous formulation of a VSR invariant quantum field theory. It is a speculation-free candidate for physics at the Planck scale. From an algebraic point of view it is a minimal departure from Lorentz symmetry while fully incorporating the conclusion that at Planck scale~\textendash~where Cohen and Glashow conjectured VSR becomes operative~\textendash~spacetime must be non-commutative~\cite{Ahluwalia:1993dd,Doplicher:1994zv}. 

The theoretical spirit of the Elko formalism is similar to the work of~\cite{SheikhJabbari:2008nc}. Both attempt a systematic construction of quantum fields that are invariant under one of the subgroups of VSR. Sheikh-Jabbari and Tureanu point out the difficulties encountered in constructing these quantum fields and it may be that those difficulties cannot be fully circumvented without following the path suggested by the Drinfel'd twist. The spin statistics theorem is built-in in the work of ref.~\cite{SheikhJabbari:2008nc}. While there is, as of yet, no formal proof of the spin statistics theorem in the Elko construct, ref.~\cite[section 7]{Ahluwalia:2004ab} essentially assures its validity. In contrast to the work of ref.~\cite{SheikhJabbari:2008nc} our formalism does not invoke the framework of twisted Poincar\'e algebras and for that reason the VSR invariant Elko fields may carry their signatures at lower energies. It remains an open problem to see if the non-commutative spacetime approach can shed new light on the locality structure of Elko.

\subsection{Darkness of the \emph{SIM}(2) invariant Elko fields with respect to the fields of the standard model} \label{sec:darknessofsim2}

In sharp contrast to mass dimension three halves associated with the fermionic matter fields of the standard model, the spin one half \emph{SIM}(2) VSR invariant Elko fields carry mass dimension one. The resulting mismatch of mass dimensions forbids the new fields to enter the fermionic doublets of the standard model. Therefore, \emph{SIM}(2) VSR invariant  Elko becomes a natural dark matter candidate.

Observational evidence suggests that dark matter is self-interacting~\cite{Spergel:1999mh}. If dark matter was fermionic and carried mass dimension three half any such self-interaction would be suppressed by two orders of the Planck scale (or, unification scale). On the other hand if dark matter is described by the here-constructed fermionic fields, then no such suppression occurs and the following self interactions
\begin{equation}
  g_\Upsilon\, \left[\dual{\Upsilon}(x) {\Upsilon}(x)\right]^2,\qquad
  g_\upsilon\, \left[\dual{\upsilon}(x) {\upsilon}(x)\right]^2
\end{equation}
with $g_\Upsilon$ and $g_\upsilon$ as dimensionless coupling constants, are dimension four operators.
The only unsuppressed coupling with the SM fields is with the Higgs 
\begin{equation}
  g_{\phi\Upsilon} \,\phi^\dag(x)\phi(x) \dual{\Upsilon}(x) {\Upsilon}(x),\qquad
  g_{\phi\upsilon} \,\phi^\dag(x)\phi(x) \dual{\upsilon}(x) {\upsilon}(x)
\end{equation}
where $g_{\phi\Upsilon}$ and $g_{\phi\upsilon}$ are dimensionless coupling constants and $\phi(x)$ represents the SM Higgs doublet. But, it is easy to do without fermionic dark matter by simply invoking a scalar (or, pseudo-scalar)  field for the purpose which, as our reader knows, have mass dimension one (like \emph{SIM}(2) VSR invariant Elko).  

However, there is one important aspect of fermionic dark matter that seems to have escaped general attention. A fermionic dark matter comes with the possiblity of supporting the dark-matter halo by Fermi degenerate pressure. Recently evidence has emerged that the Milky way is embedded in a spheroidal distribution of dark matter~\cite{Law:2009yq}.\footnote{The text that follows in this paragraph needs to be revised as it fails to take into account the SM-matter content. It is left here unedited to mirror the published version.} 
 Assuming the spheroid consists of Elko dark matter, the standard arguments of balancing the Fermi pressure against the gravitational potential energy yield the following relationship between the Elko mass and Chandrasekhar value for the halo's size~\cite[eq. 10.18]{Ahluwalia:2004ab}
\begin{equation}
  R_{\mathrm{Ch}} \sim {x^{-2}_{\mathrm{Elko}}} \; 6.3 \times 10^{-2}\; \mathrm{pc}
\end{equation}
where $x_\mathrm{Elko}$ represents the Elko mass $m$ expressed in $\mathrm{keV}$. Taking the radial dimensions of the halo to be about $60\;\mathrm{kpc}$, we immediately infer $m \sim 1\; \mathrm{eV}$. As such if dark matter is indeed Elko, then an eV mass range brings it intriguingly close to the sterile neutrino mass hinted at by the LSND and MiniBooNE experiments~\cite{AguilarArevalo:2010wv}.\footnote{Needless to say that this last argument depends only on the fermionic aspect of Elko and as such it applies to all fermionic dark matter candidates.} Arguments in favour of an eV range sterile neutrino are also supported by the analyses found in ref.~\cite{Nieuwenhuizen:2008pf,Giunti:2010zu}.
 
\subsection{Elko gravity}
 
Since the symmetries of the  event space underlying \emph{SIM}(2) VSR invariant Elko and those associated with the standard model are no longer identical, it is not clear to the authors if the gravitational coupling between (a) the standard model particles, (b) Elko particles with the standard model particles, and (c) Elko particles with Elko particles, are identical. Even the most naive treatments that couple Elko to the general-relativistic gravity, which we interpret as the theory of gravity for the SM matter and gauge fields (or, at least the fields that carry undeformed Poincar\'e symmetries),  exhibit significant differences~\cite{Boehmer:2007dh,Shankaranarayanan:2009sz,Wei:2010ad,Boehmer:2006qq,Gredat:2008qf}. From the perspective of our work, the key difference between Elko and SM  fermions lies in the relative helicity structure between a-type ($r$-type, for SM)  and b-type ($\ell$-type, for SM) components of the spinors. It is opposite for the former while the same for the latter ($\wp\, \Theta\left[\chi^b(p^\mu)\right]^\ast$ carries opposite helicity to that of $\chi^b(p^\mu)$). If the spin and angular momentum content of the Elko and the SM matter fields is ignored the above-mentioned scenarios are expected to be gravitationally equivalent (and can thus be approximately described  by the theory of general relativity). The difference, we expect, arises when spin and angular momentum become important. This also becomes apparent from a series of works on the subject~\cite{Gibbons:2007iu,Chaichian:2008pq,Bogoslovsky:1998wa,Li:2010zb,Silagadze:2010bi,Das:2010cn,Vacaru:2010fi,Calcada:2002sw,Cohen:2006sc,Kouretsis:2008ha}.

\section{Conclusion}

The null result of the Michelson-Morely experiment, combined with the demand of \emph{P, T, CP}, or \emph{CT} invariance (or, quantum field theoretic locality), makes the Lorentz group a necessity. Otherwise, the \emph{HOM}(2) and the \emph{SIM}(2) subgroups of very special relativity suffice to preserve many of the essential results of the theory of special relativity. Here we abandoned the invariance under the indicated discrete symmetries, and found that the damage to the quantum field theoretic locality resides entirely in a single additive term to the field-momentum anticommutator. Even that term identically vanishes when one confines one's attention to the VSR preferred direction. 

This `damage' to locality, however, produces the startling result that one now obtains a mass dimension one fermionic field for spin one half. This has dramatic consequences for self-interactions of the new fields, and their interactions with the fields of the standard model of high energy physics (see section\;\ref{Sec:qf} and section\;\ref{sec:darknessofsim2}). In fact one ends up obtaining a quantum field theoretic structure that seems most suited to describe dark matter. Given that the theory of special relativity may be considered as a theory that reflects symmetries associated with rods and clocks made of the standard model fields, it is tempting to suggest that very special relativity does the same for the dark matter rods and clocks. This, we think, is the essential challenge that this communication raises for the theoretical physics community. 

\section*{Acknowledgements}

One of us (DVA) thanks Harvey Brown for an extended discussion on the view of relativity taken here~\cite{Brown:2005hr}. We are also thankful to Cheng-Yang Lee for our continuing discussions on the Elko-VSR issues. Both authors are grateful to the facilities of the Kaikoura and the Westport Field Stations of the University of Canterbury where much of this work was done.

\providecommand{\href}[2]{#2}\begingroup\raggedright\endgroup

\end{document}